\documentclass[a4paper,11pt]{article}
\usepackage{pos}

\title{The soup that is not too hot}
\author*[a,b]{Maria Stefaniak}
\affiliation[a]{GSI Helmholtz Centre for Heavy Ion Research,\\ Planckstraße 1, 64291 Darmstadt, Germany}
\affiliation[b]{Warsaw University of Technology\\ pl. Politechniki 1, 00-661 Warszawa, Poland}
\emailAdd{m.stefaniak@gsi.de}
\emailAdd{maria.stefaniak@pw.edu.pl}

\abstract{The exploration of the QCD phase diagram is the goal of many experiments in the world. The low net baryon density region is relatively well studied, while the one corresponding to higher net densities still contains multiple unrevealed puzzles getting significant attention. In multiple experimental complexes such as STAR or HADES, the heavy ions collide at relatively lower energies than those obtained at LHC, creating "not the hottest soup" of nuclear matter. The examination of it provides unique insights into understanding the properties of matter production and its transitions. In this paper, several of the significant experimental discoveries are reviewed.}

\FullConference{%
  FAIR next generation scientists - 7th Edition Workshop (FAIRness2022)\\
  23-27 May 2022\\
  Paralia (Pieria, Greece)
}

\begin{document}
\maketitle

\section{Introduction}

\begin{figure*}[h!]
\centering
\includegraphics[width=0.9\textwidth]{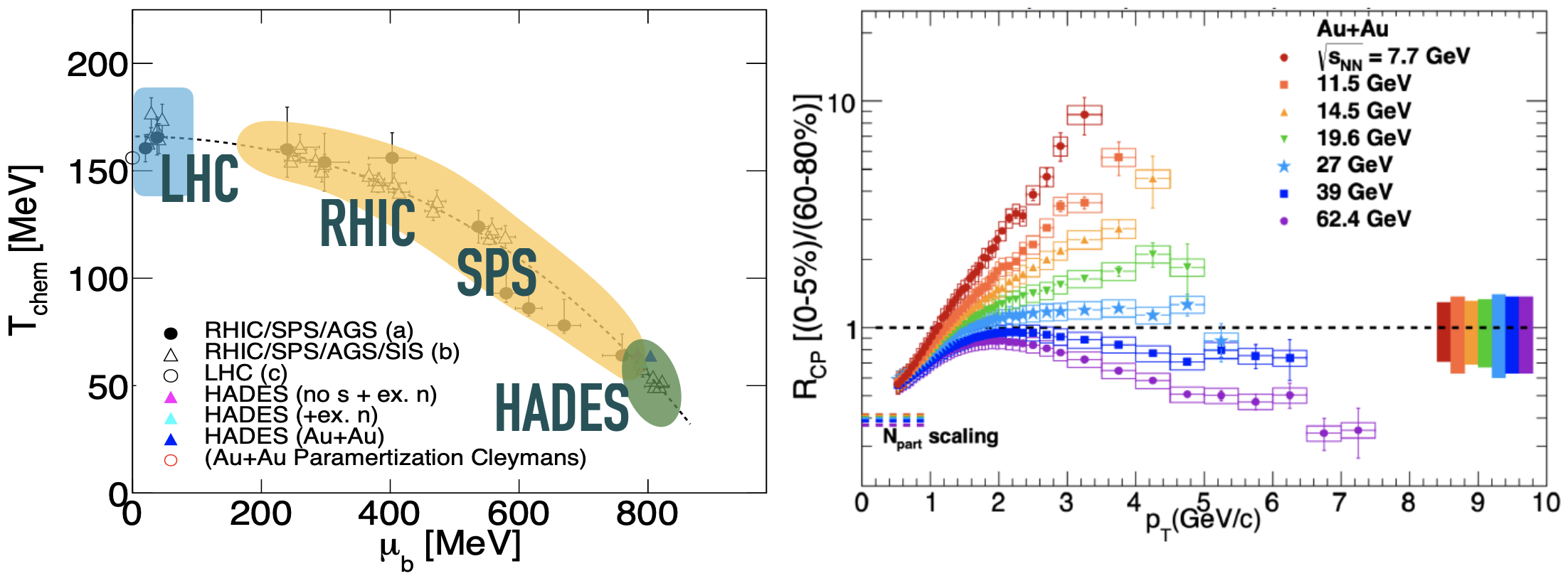}
%\decoRule
\caption{a) QCD phase diagram, with depicted regions of T and $\mu_B$ corresponding to these characterizing the matter created at given experiments and facilities. b) Charged hadron $R_{cp}$ for RHIC BES energies \cite{STAR:2017ieb}.}
\label{fig1}
\end{figure*}

Studies of the properties of the strongly interacting matter are the goals of many experiments worldwide. The investigation of the so-called "hottest soup" - Quark-Gluon Plasma (QGP) - attracted nearly the full attention of heavy-ion physicists. Recently, the relatively colder and "denser" regions of the QCD phase diagram (shown in Fig. \ref{fig1}a) gained the interest of the community, still having many unrevealed puzzles such as transitions between partonic and hadronic states or description of Neutron Stars (NS) and Neutron Star Mergers (NSM) Equation of State (EoS). This paper briefly summarizes the most vital experimental discoveries in that diagram region.

\section{Onset of QGP}

%The QGP phase of the matter is expected to be at higher temperatures, where the energies of collided ions are large enough for the deconfinement of partons. However, it is still not precisely determined at which conditions we expect the presence of the QGP and when that state is not obtained. 
\paragraph{Nuclear modification factor:}
The non-monotonic collision energy dependence of nuclear modification factor $R_{cp}$ is one of the possible signatures of the disappearance of the partonic degrees of freedom. It is described with the following equation:
\begin{equation}
    R_{CP} = \frac{\langle N^{per}_{bin}\rangle d^3N^{cen}_{AA}/d\eta d^2p_T}{\langle N^{cen}_{bin} \rangle d^3N^{per}_{AA}/d\eta d^2p_T}
    \label{R_cp}
\end{equation}
The most peripheral AA collisions are expected to mimic the NN collisions and are used in the following studies due to the lack of data for proton-proton collisions at such a big range of collision energies. When the values of $R_{CP}$ are lower than unity, the matter expansion is suppressed. The opacity of the partonic deconfined medium explains it. On the other hand, the enhanced $R_{CP}$ is expected to be the effect of the domination of the hadronic interactions. The results obtained at STAR experiment \cite{STAR:2017ieb} are shown in the Fig. \ref{fig1}b. The clear $\sqrt{s_{NN}}$ dependence is visible. The suppression of $R_{CP}$ is treated as a signature of the presence of the QGP phase for the higher energies. 

\begin{figure*}
\centering
\includegraphics[width=0.7\textwidth]{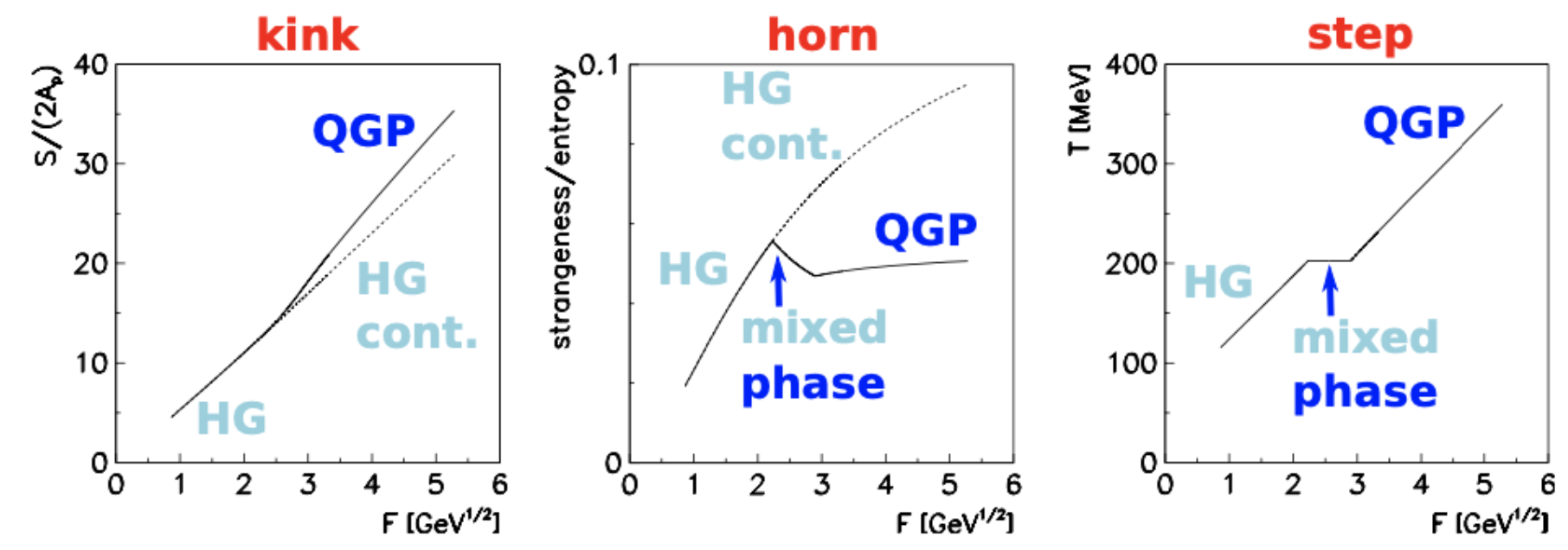}
%\decoRule
\caption{The kink, horn, and step structure proposed in \cite{Gazdzicki:1998vd} with depicted expected phases: HG - hadronic gas, QGP, and mixed phase.}
\label{fig:kink}
\end{figure*}

\begin{figure*}
\centering
\includegraphics[width=0.9\textwidth]{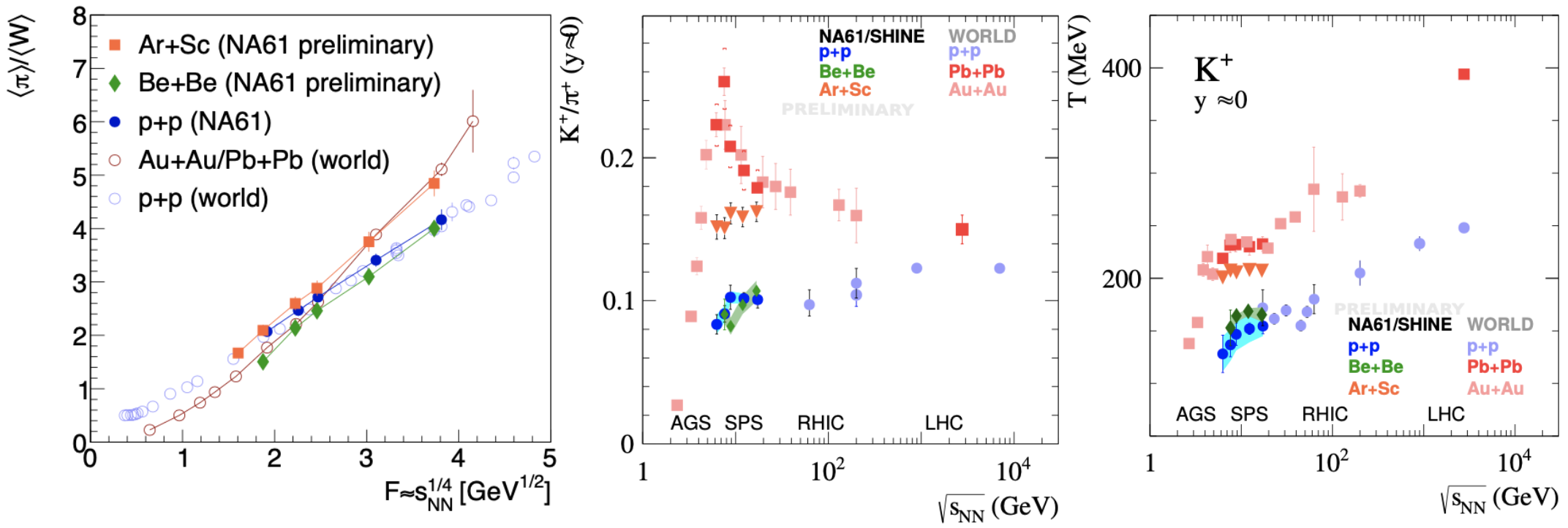}
%\decoRule
\caption{Energy dependence of: (from left) the total $\pi$ multiplicity divided by the number of wounded nucleons W, $K^+$ to $\pi^+$ multiplicity ratio and temperature extracted form K $p_T$ spectra \cite{Aduszkiewicz:2017mei}.}
\label{fig:kink_exp}
\end{figure*}

\paragraph{Kink, horn, step:}
In \cite{Gazdzicki:1998vd} authors suggested that according to the Statistical Model of the Early Stage (SMES), where a number of heavy flavors and whole entropy are the same before and after hadronization, measurements of entropy, strangeness, and temperature of the system in the function of Fermi energy are vital for the investigation of the onset of QGP. The changes in slopes of three quantities are explained as the activation of the partonic degrees of freedom, which is sketched in Fig. \ref{fig:kink}. The NA61/SHINE collaboration used pions' production as a proxy for the entropy, charged kaons to pions ratio to stand for strangness to entropy, finally, and the temperature extracted from the inverse slope parameter of transverse mass spectra. Obtained results (see Fig. \ref{fig:kink_exp}) are in remarkable agreement with the theoretical predictions. The slope changes are present for the AA collisions, while, in the case of p+p, where the QGP phase is not expected, are negligible. 

\section{First order phase transition}

\begin{figure*}
\centering
\includegraphics[width=0.8\textwidth]{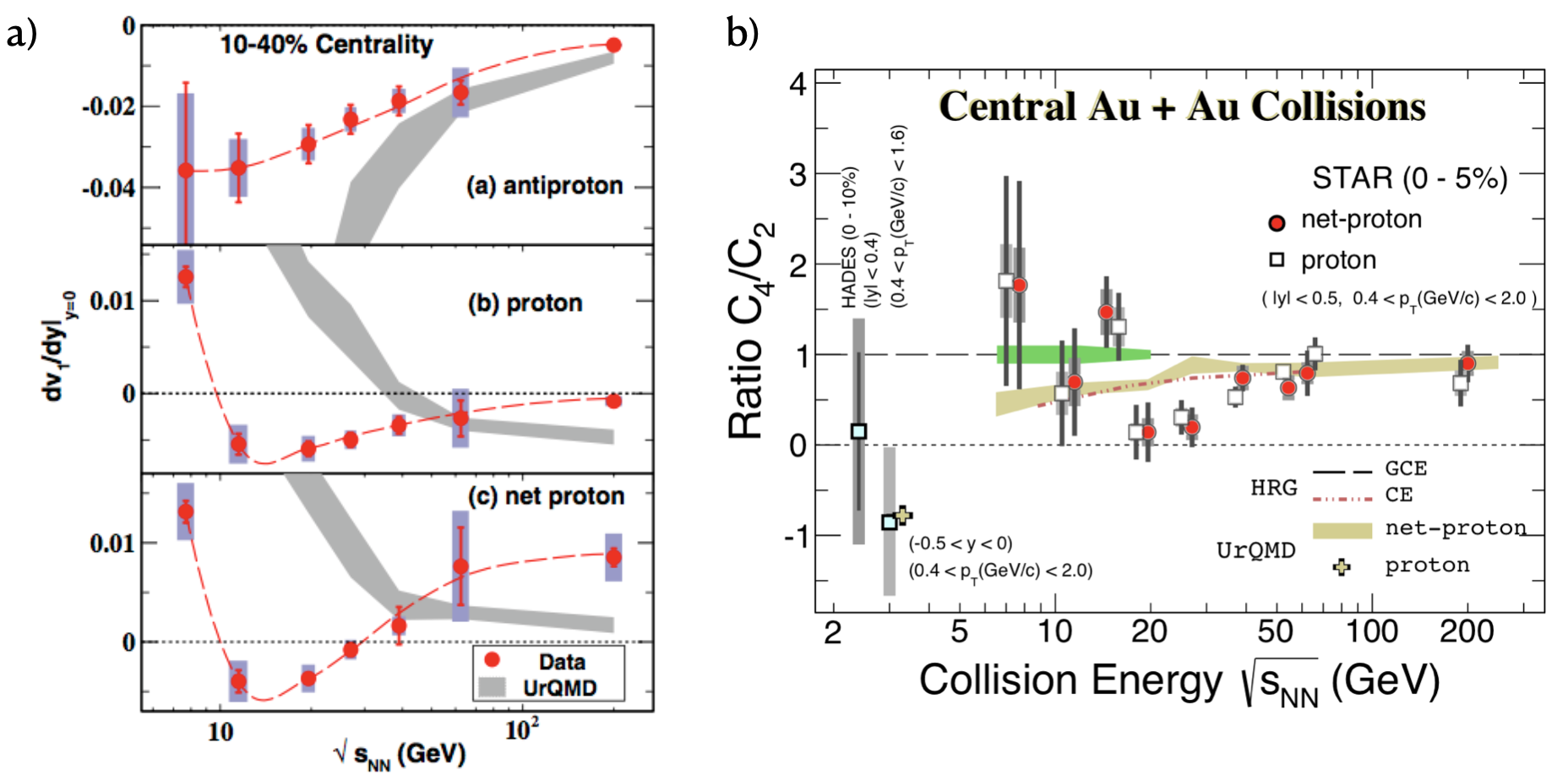}
%\decoRule
\caption{a)Directed flow slope $dv_1/dy$ of net-protons near midrapidity versus beam energy for intermediate-centrality ($10-40\%$) Au+Au collisions: STAR experiment data and UrQMD simulation. \cite{STAR:2014clz}b) Collision energy dependence of the ratios of cumulants, $C_4/C_2$, for proton (squares) and net-proton (red circles) from various experiments \cite{STAR:2021fge}.}
\label{fig:cumulants_exp}
\end{figure*}

\paragraph{Directed flow of protons:} $v_1$ is sensitive to the early stage of collision and compression of the medium. The $v_1$ slope in function of rapidity for all the hadrons does not change with the energy of the collision, except of protons \cite{STAR:2014clz}. A system undergoing a $1^{st}$ order phase transition characterized by the small pressure gradient due to mix phase formation causes the change of sign in the slope of $dv_1/dy$ for net baryons and/or due to softening of the EoS \cite{Ohnishi:2017xjg}. 
STAR performed the detailed measurements of $v_1$ for various collisions energies and the non-monotonic trends are visible (Fig. \ref{fig:cumulants_exp}a). It is treated as a signature of the first order phase transition and justified additionally by lack of agreement with the UrQMD simulations \cite{Bleicher:1999xi}, where the transition is not included. 

\section{Critical Point}
\paragraph{Fluctuations of net-baryon distributions:}

The non-monotonic behavior in the event-by-event fluctuations of globally conserved quantities is treated as one of the signatures of the presence of the Critical Point (CP) \cite{Sombun:2017bxi,Luo:2017faz}. % Cumulants used for these measurements can be expressed with:
%\begin{align*}
%    \delta N &=N-<N>  \\
%    C_1 &= <N>  \\
%    C_2 &= <(\delta N)^2> = \sigma^2 \\
%    C_3 &= <(\delta N)^3> \\
%    C_4 &= <(\delta N)^4> - 3 <(\delta N)^2>^2   \\
%    \kappa \sigma^2 &= \frac{C_4}{C_2}
%    \label{eq:cumulants}
%\end{align*}
The higher the cumulant order, the more sensitive it is to the correlation length. $4^{th}$ order is predicted predicted to have a non-monotonic energy dependence due to contribution from the CP. The observed in \cite{STAR:2021fge} $C_4/C_2$ suppression consistent with fluctuations driven by baryon number conservation indicates the domination of hadronic interaction - visible for lower $\sqrt{s_{NN}}$ in Fig. \ref{fig:cumulants_exp}b.

\section{Neutron stars and neutron star mergers}
\paragraph{Virtual photons:}

\begin{figure*}
\centering
\includegraphics[width=0.7\textwidth]{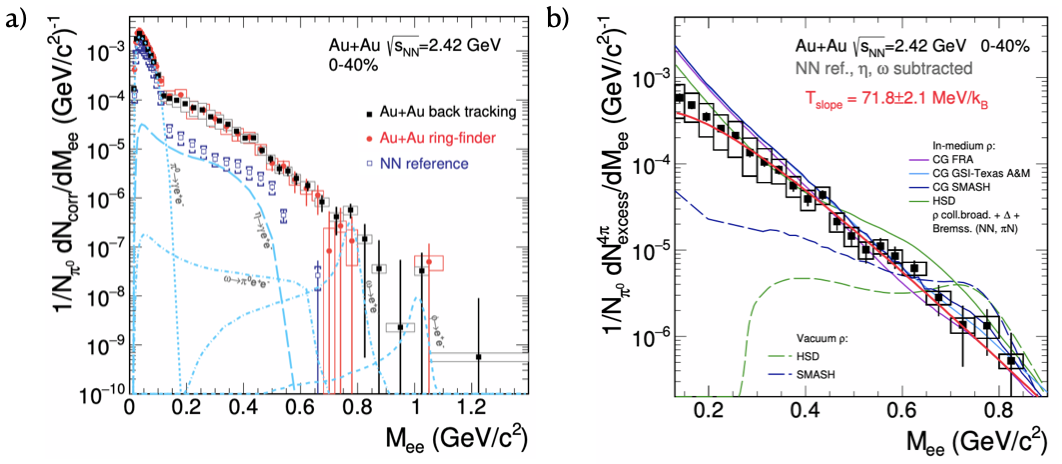}
%\decoRule
\caption{a) Reconstructed $e^+e^-$ mass invariant distribution from Au+Au collisions and expected contributions from mesonic decays. b) Acceptance corrected dilepton excess yield \cite{HADES:2019auv}.}
\label{fig:hades}
\end{figure*}

Moving further on the phase diagram towards the higher baryon densities and lower temperatures, there is a region corresponding to NS ($T<10$ MeV, $\rho<10 \rho_0$\footnote{$\rho_0 \approx 0.16 fm^{-3}$ - density of normal matter},) and NSM ($T<50$ MeV, $\rho<2-6 \rho_0$). The systems with similar properties ($T: 80-100$ MeV, $\rho<2-3 \rho_0$) can be created in the HADES experiment, what can serve as a reference for the astrophysical research of NS and NSM. The production of virtual photons, which are expected to be created during the whole evolution of the system's evolution can provide the information about the temperature of the system and consequently contribute to the process of EoS determination. In the Fig. \ref{fig:hades}a, HADES used the reconstructed $e^+e^-$ mass distribution as a manifestation of virtual photons \cite{HADES:2019auv}. %The data were corrected for the acceptance in dilepton excess yield via subtraction of $\eta$, $\omega$ contributions, and NN reference normalization to the number of neutral pions. 
The obtained excess radiation (Fig. \ref{fig:hades}b) shows nearly-exponential fall-off. The extracted temperature of the system's evolution is equal to $T = 71.8\pm 2.1$ MeV/$k_B$, while theoretical models predict the $T=50-80 $ MeV of the post-merger NS around the dense remnant core.

\section{Conclusions}
Many observables were investigated, providing an excellent reference for the theoretical studies of properties of the strongly interacting matter and its transitions. The impressive spectrum of data collected by STAR, NA61/SHINE, and HADES experiments is still under investigation, possibly leading to ground-breaking discoveries and establishing the EoS valid for the whole QCD phase diagram.

\section{Acknowledgments}
I want to acknowledge the financial support of IDUB-POB-FWEiTE-3 project granted by Warsaw University of Technology under the program Excellence Initiative: Research University (ID-UB). This work is supported by the scholarship grant from the GET$\_$INvolved Programme of FAIR/GSI.

%\bibliographystyle{unsrt}
%\bibliography{reference.bib}% Produces the bibliography via BibTeX.

\begin{thebibliography}{99}
\bibitem{STAR:2017ieb}
L. Adamczyk et al.: Phys. Rev. Lett., 121(3):032301, 2018
\bibitem{Gazdzicki:1998vd}
Marek Gazdzicki and Mark I. Gorenstein: Acta Phys. Polon. B, 30:2705, 1999
\bibitem{Aduszkiewicz:2017mei}
Antoni Aduszkiewicz: Nucl. Phys. A, 967:35–42, 2017
\bibitem{STAR:2014clz}
Adamczyk et al.: Phys. Rev. Lett., 112(16):162301, 2014
\bibitem{Ohnishi:2017xjg}
A. Ohnishi, Y. Nara, H. Niemi, and H. Stoecker: Acta Phys. Polon. Supp., 10:699, 2017
\bibitem{Bleicher:1999xi}
M. Bleicher et al.: J. Phys. G, 25:1859–1896, 1999
\bibitem{STAR:2021fge}
M. S. Abdallah et al.:  Phys. Rev. Lett., 128(20):202303, 2022
\bibitem{Sombun:2017bxi}
S. Sombun et al.: J. Phys. G, 45(2):025101, 2018
\bibitem{Luo:2017faz}
Xiaofeng Luo and Nu Xu: Nucl. Sci. Tech.,28(8):112, 2017
\bibitem{HADES:2019auv}
J. Adamczewski-Musch et al.: Phys., 15(10):1040–1045, 2019





\end{thebibliography}

\end{document}